# Search for explanation of the neutron lifetime anomaly

A. P. Serebrov,* M. E. Chaikovskii, G. N. Klyushnikov, O. M. Zherebtsov, A. V. Chechkin.
*Petersburg Nuclear Physics Institute, NRC "Kurchatov Institute",*
*188300 Gatchina, Leningrad District, Russia*

Abstract

All results of the neutron lifetime measurements performed in the last 20 years with the UCN storage method are in a good agreement. These results are also consistent with the latest most accurate measurements of the neutron decay asymmetry within the Standard Model. However, there is a significant discrepancy at 3.6σ (1% of the decay probability) level between the averaged result of the storage method experiments and the most accurate measurements performed with the beam method. This article addresses the possible causes of the neutron lifetime discrepancy. We focused on finding the spectrum of possible systematic corrections in the beam experiment. Four separate sources of the systematic errors which had not been properly addressed in articles dedicated to the beam technique were considered. Two of those sources are related with the motion of protons in an electromagnetic field and the elastic scattering by the residual gas. These problems are associated with the geometrical configuration of the setup and the proton detector size. The Monte-Carlo simulation shows that these effects are negligible and do not affect measurement results. The third error concerns proton loss in the dead layer of the detector. It is shown that the correction for the dead layer requires a more detailed analysis than that given in the papers describing the beam measurement method. The fourth source of the systematic error is the charge exchange process on the residual gas. The influence of the residual gas was neglected in the beam method experiments. We present arguments that careful analysis of this issue is necessary since the proposed proton losses correction decreases the measured lifetime bringing it closer to the storage method results. For the problem of proton charge exchange on a residual gas, the spectrum of possible corrections is investigated. It is shown that for an accurate calculation of the correction, it is necessary to directly measure the concentration and composition of the residual gas inside the proton trap. The analysis reveals that even presence of only $H_2$ molecules inside the proton trap can lead to the significant correction which is the most probable explanation of the neutron anomaly.

A. P. Serebrov,* E-mail: serebrov_ap@pnpi.nrcki.ru

## I. **INTRODUCTION**

The neutron is the most long-living of all unstable elementary particles with a lifetime of about 880 seconds. The first experiments to measure the neutron lifetime were conducted more than 70 years ago, yet, even now new researches are carried out and measurement techniques are being improved. Such interest to the topic can be explained by an importance of precise measurements of the neutron lifetime for elementary particle physics and cosmology. However, the relatively high value of the neutron lifetime creates certain difficulties in performing measurements with an accuracy required to validate existing theoretical models. Currently, the accuracy of measurements is of order 0.1%; however, there exists an unresolved problem of the 1% disagreement between the neutron lifetime results obtained with two main methods.

In the Standard Model (SM) of elementary particles the neutron decays into the proton, electron and electron antineutrino. There is also a small probability of decay with an additional photon or with a hydrogen atom in the final state:

$$
\begin{aligned}
n &\to p + e^- + \bar{\nu}_e & &100\% \\
n &\to p + e^- + \bar{\nu}_e + \gamma & &(9.2 \pm 0.7) \cdot 10^{-3} \\
n &\to H + \bar{\nu}_e & &3.9 \cdot 10^{-6}
\end{aligned} \quad (1)
$$

There are two fundamentally different approaches to measure the neutron lifetime. The first approach is based on detecting the products of β-decay (protons or electrons) during the period when a neutron beam passes an experimental apparatus. In the second approach, one studies how the amount of neutrons in an experimental volume changes with time. Strictly speaking, in these two methods different physical constants are measured: if one detects decay products then only the β-decay probability is measured, while the second method gives an opportunity to use the neutron counts to measure the total decay probability regardless of a decay channel and final state particles. Even within the SM these two probabilities are distinct. The difference of $4 \cdot 10^{-4}$ % is due to the decay with a hydrogen atom in the final state, but this small fraction is usually neglected. To declare that both methods can be used to measure the neutron lifetime means to assume that the β-decay (1) is the only possible neutron decay channel.

Registration of the decay products is performed in the so-called "beam" method of β-decay probability measurements. In this method a beam of cold neutrons passes through a vacuum system in which electrical field is applied to guide decay products onto particle detectors. In later implementations of this method a neutron beam passes through the electromagnetic proton trap. Protons emitted in neutron β-decays are trapped by the field and hence can be stored in a trap for a certain time period. After the storing period the protons are guided to the detector (Fig. 1). The implementation of the proton trap decreased the experimental uncertainties to ≈3s [1].

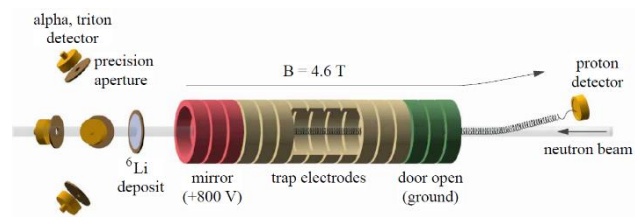

FIG. 1. Experimental apparatus used in the last most precise beam measurement of the neutron lifetime.

To perform experiments using the beam method precise measurements of an initial neutron flux and proton counts are required, because the final result is derived on the basis of the ratio of these two parameters.

The measurements with neutron counting are carried out using the storage of ultracold neutrons (UCN). A feature of UCNs is that they can be stored in magnetic or material traps for a long period of time because the UCNs reflect from the trap walls. There is a probability to lose a neutron in a collision with a trap wall, but it can be decreased to very low values. In the modern cryogenic material traps one can reach 1-2% of the neutron decay probability [2, 3], and even lower losses can be achieved with magnetic traps [4, 5, 6, 7].

The experimental technique is to fill UCNs into the trap and count the remaining neutrons after a certain period of time. Then this procedure should be repeated with another period of time in order to calculate the neutron lifetime in the trap. This value differs from the neutron lifetime by the correction based on the neutron loss probability in collisions with the trap walls. Since that correction is small the UCNs provide a way to measure the neutron lifetime almost directly.

In order to take into account this correction, the measurements are usually carried out for several geometric configurations of the trap and/or in various energy ranges. The final value of the neutron lifetime is then calculated using the extrapolation to zero loss probability. In experiments with UCN traps one measures the total neutron lifetime, denoted as $\tau_n$, or total neutron decay probability regardless the decay products in the final state.

The problem of the disagreement between two described methods appeared for the first time in 2005, when the result $\tau_n = 878.5 \pm 0.7 \pm 0.3\ s$ [2] was obtained using a cryogenic UCN trap with the gravitational locking of neutrons in the trap. For that moment the value of the neutron lifetime according to PDG was $885.7 \pm 0.8\ s$ and the result $\tau_\beta = 886.6 \pm 1.2_{stat} \pm 3.2_{sys}\ s$ obtained shortly before using the beam method [1] was included into the derivation of the PDG value, and it confirmed the earlier results (see Fig. 2). Therefore, the result of the new experiment based on UCN storage deviated from the PDG value by 6σ and has significant disagreement with the mentioned beam experiment result. However, further experiments with implementation of magnetic and material traps confirmed the discrepancy [8] Furthermore, most UCN results have been revised after work [9], which led to appearance of the so-called "neutron anomaly" [10]. Currently, the conventional value of the neutron lifetime according to PDG is $879.4 \pm 0.6\ s$ and it is determined by the results of only the UCN storage experiments. Yet the results of the beam method experiments still have significant disagreement with that value.

In the last few years several experiments with UCN storage in magnetic and material traps were carried out. The results are shown in Fig. 2, the discrepancies between the new UCN experiments do not exceed two standard deviations. However, since the anomaly appeared there have been no new beam experiments (the result of beam experiment in 2013 [11] was calculated after additional analysis of the data obtained in 2004). And finally, the latest results of the β-decay asymmetry measurements are in agreement with neutron lifetime obtained in experiments with UCN storage [12] and CKM unitarity [13].

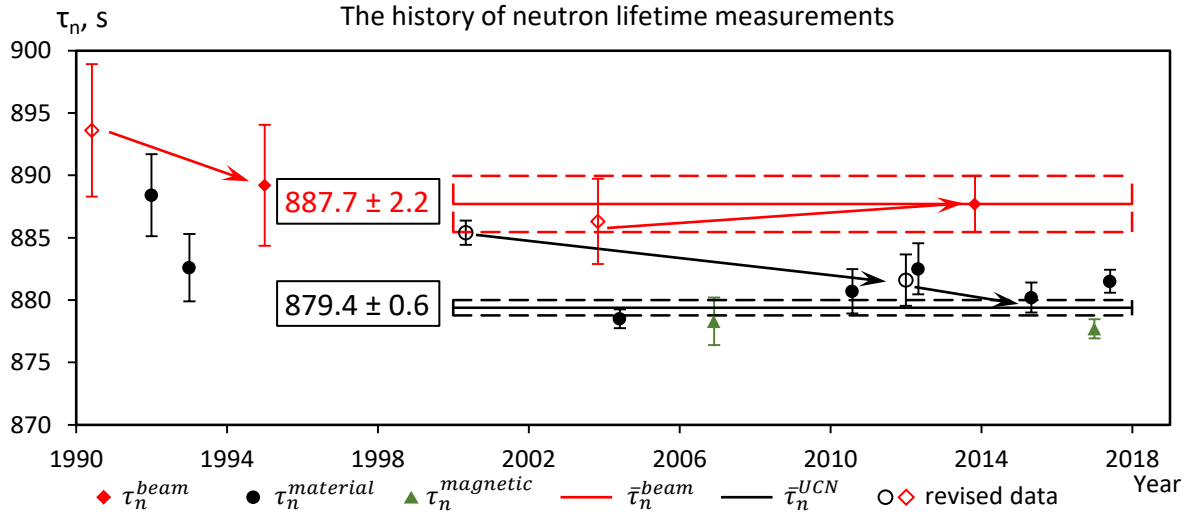

FIG. 2. History of the neutron lifetime measurements since 1990. Arrows represent experiment revisions. Storage experiments have been revised after work [9], and beam experiment revision is presented in [11].

For the moment, the reason for the neutron anomaly is not discovered. One can consider two approaches to solve this problem either by searching for deviations from SM i.e. search for some new physics to explain the discrepancy between the neutron lifetime and the β-decay period, or by providing an analysis of additional systematic errors which were overlooked. Additional errors in the UCN storage experiments and various explanations of neutron anomaly with processes beyond SM were considered in ref. [14]. The resulting conclusion of the provided analysis is that the explanation of the anomaly lies outside that area.

For the reasons listed above, we assume that one should search for a solution of the neutron anomaly problem in systematic uncertainties of the beam method experiments. Currently the main experiment cited when talking about beam measurement result is from the article [1]. Therefore, we want thoroughly examine possible sources of the systematic errors in this experiment. In that article authors present the detailed analysis of several systematic errors regarding the neutron flux. But here we would like to discuss problems which might occur during protons movement in the trap and their registration.

Recently, the experimental efforts were focused on the UCN storage experiments and no new beam method results were published. Therefore, to solve the neutron anomaly problem it is essential to perform a new more accurate beam experiment. Such experiments are currently under preparation [15, 16], therefore, now it is very important to consider all possible systematic uncertainties, especially those overlooked in the previous experiments.

## II. THE SYSTEMATIC ERROR ASSOCIATED WITH THE DETECTOR DEAD LAYER

Our analysis of the systematic errors of the beam measurement method is based on the description of the experimental setup used to obtain the most accurate result using this method. In ref. [1] the detailed description of the experimental setup is presented along with the obtained data and the analysis of some systematic errors. Further analysis of the systematic errors is based entirely on the data presented in ref. [1].

The employed experimental technique relies on the ability to accurately count the protons emitted in β-decay. The neutron lifetime is calculated based on ratio of the proton counts to the amount of events registered by the neutron monitor, hence any additional proton losses lead to a measured lifetime which is higher than the real value. The proton trap length and hence the number of decays were varied and the experimental results were obtained using the derivation of the slope of the curve which describes the dependence of decay number on the trap length. Therefore, the losses of a proton fraction lead to the proportional systematic error.

The first step in the simulation that considers the loss of protons is to calculate the intensity of the protons emitted from the trap as a function of the distance from the center of the trap and check that all the protons arrive at the detector.

The paper [1] has two graphs of the neutron beam as a function of the beam radius. One shows the point in the end of the proton trap and the other one shows the point of the neutron monitor. We noticed that the authors show only the neutron beam radius but do not take into consideration widening of the proton trajectories due to their interaction with the magnetic field. Besides, in order to correctly estimate proton loss one should know their distribution over detector surface. In order to simulate the initial proton beam from the trap we have to know the intensity of the neutron beam in the trap as a function of the radius. Unfortunately, authors of work [1] show the fitting function for the intensity only for the position of the neutron monitor:

$$\varphi(x) = e^{-\frac{x^b}{2c^2}}, \quad (2)$$

where $b = 2.59$, $c = 9.73$ (left of Fig. 3). In order to estimate the intensity of the beam in the trap we used data from the figure 11 of the paper [1]. We assumed that the intensity as a function of the radius has the same form inside the trap as it has at the neutron monitor, but with another value of parameter $c$. This parameter was adjusted in the way that the measured and fitted neutron distributions at the end of the trap were as close as possible (Fig. 4.b).

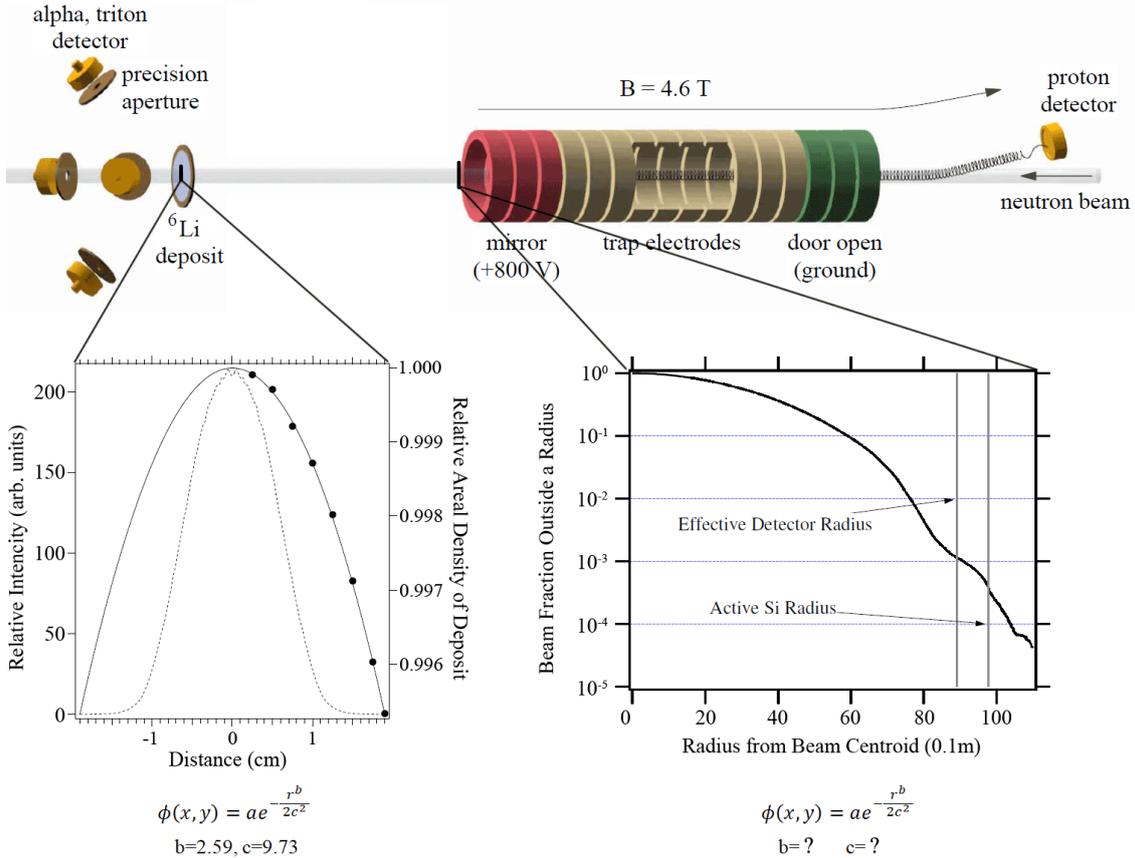

FIG. 3 Composite schematic illustration of the beam intensity measurements in the Nico et. al [1] experiment based on their data of the neutron distributions in two points of the neutron beam.

The obtained function fits the data very well and the value of fitting parameter $c = 5.05$. This value indicates that the neutron beam width increases on its way from the trap to the monitor by a factor of $\left(\frac{9.73}{5.05}\right)^{2/b} \sim 1.66$. We used the function (2) with coefficients $b = 2.59$, $c = 5.05$ in our further analysis of the proton beam.

The Fig. 5a illustrates the results of the simulation of neutron decay points in assumption that the probability to find neutron in a circle of radius r is:

$$F(r) = \frac{\int_0^r x\varphi(x)dx}{\int_0^R x\varphi(x)dx}, \quad (3)$$

where $\varphi(x)$ is the function (2) with parameters $b = 2.59$, $c = 5.05$.

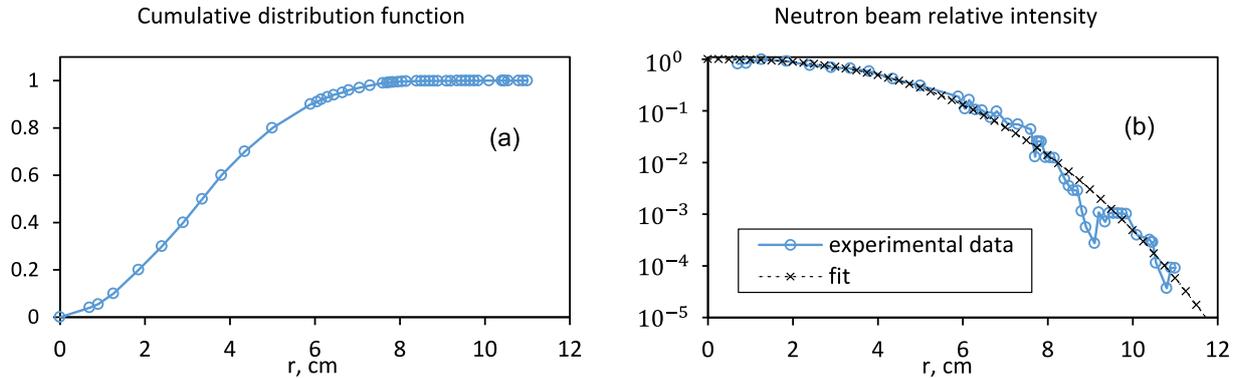

FIG. 4 The distribution function (a) and the relative beam intensity of the neutrons (b).

The Fig. 5b shows the points where the protons, with initial positions in the trap illustrated in Fig. 5a, hit the detector. The presented result includes $10^5$ proton trajectories. The simulation revealed that for the neutrons all decay points have radial coordinate less than 11 mm (Fig. 5a) and for the protons detector hit points have radial coordinates less than 12 mm (Fig. 5b). The radial coordinates exceed 10 mm for 30 neutrons and 33 protons. Therefore, we conclude that amount of initial protons outside the detector radius (see Fig. 1) is at level $3 \cdot 10^{-4}$. Only 3 of 100000 protons had radial coordinate at the detector over 11 mm.

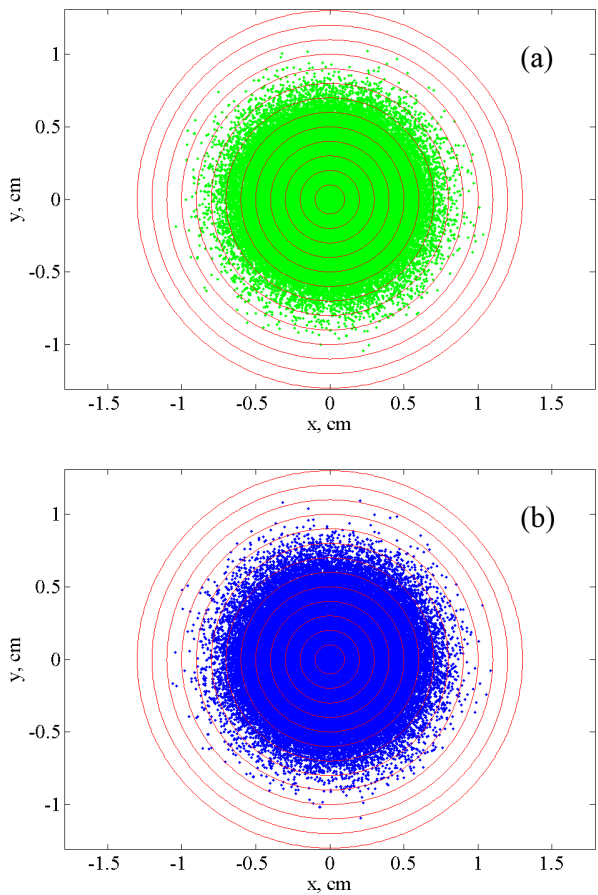

FIG. 5. The results of simulations of the neutron decays (a) and protons hits (b) of the detector in a model with homogenous electromagnetic field.

From this simulation we make a conclusion that, even if we take into account the broadening of the proton beam relatively to the neutron beam, the losses of the protons related to finite size of the detector are negligible in comparison with statistical uncertainties of the experiment. The calculated proton distribution is used in our further simulations.

The second problem that we want to consider is the possible loss of protons while registering with a semiconductor detector. When registering protons, two sources of losses arise: backscattering and stopping of protons in the dead layer. Both problems are exacerbated by the need to set a lower registration threshold to reduce the background.

The problem of proton losses requires the simulation of the process of their passage through the dead layer and the calculation of the ionization energy left in the active layer of the detector. As a result, it is possible to obtain the relationship between the registered, scattered back and lost in the dead layer protons. And such modeling was carried out quite thoroughly by the authors of the work. [1].

In the experiment under consideration one cannot count all backscattered protons as completely lost since the electromagnetic field guides them back to the detector. To correctly take into account the loss of protons, it is necessary to determine what fraction of backscattered protons will actually not be registered. To solve this problem, Nico et al. [1] carried out measurements with three different values of the accelerating potential and for four detectors with different thicknesses of the dead layer. In three detectors, the dead layer consisted of 20, 40, and $60\,\mu g/cm^2$ of gold, respectively. The dead layer of the fourth detector consisted of silicon oxide. Different types of the dead layer are characterized by a different proportion of lost and backscattered protons. These values were calculated for each layer+energy combination for which the neutron lifetime was measured. The detailed results are given in work [1] and the main results are also presented in Table I. The thicknesses of the gold and silicon oxide layers used for the simulation were determined from the measured energy loss in the dead layer.

In total, the measurements were made for 9 energy+layer combinations. For each combination, Table I shows the type of the dead layer, the value of the accelerating potential and the fraction of backscattered and lost protons.

Table I Backscattered and lost protons according to the results of work [1].

| Gold layer µg/cm² | E, keV | $f_{Bsc}$ (%) | $f_{Lost}$ (%) |
|---|---|---|---|
| 20 | 27.5 | 0.664 | 0.006 |
| 20 | 30 | 0.578 | 0.006 |
| 20 | 32.5 | 0.477 | 0.004 |
| 40 | 27.5 | 1.490 | 0.138 |
| 60 | 27.5 | 2.374 | 0.799 |
| 60 | 32.5 | 1.819 | 0.414 |
| SiO$_2$ | 27.5 | 0.242 | 0.266 |
| SiO$_2$ | 30 | 0.194 | 0.201 |
| SiO$_2$ | 32.5 | 0.183 | 0.151 |

To solve the problem of uncertainty in the fraction of losses among backscattered protons, the authors performed a linear extrapolation to zero backscattering. The result of this extrapolation is the final value for the neutron lifetime presented in [1].

We want to take a closer look at the validity of using linear extrapolation to get the result and expand the modeling, taking more parameters into account. The statement about the applicability of the linear extrapolation is equivalent to the hypothesis that for all types of the dead layers and for all energies the probability that a backscattered proton will be turned around by the field and registered is approximately the same.

This statement seems to us insufficiently motivated. To simulate the passage of protons through the dead layer in the considered experiment, the SRIM 2003 program was used [17]. We carried out a similar simulation, but using a newer version of the SRIM 2013 program, and compared the spectra of backscattered protons for different layer+energy combinations. Fig. 6 shows two such spectra for the gold layers of 20 and $60\ \mu g/cm^2$ at the same accelerating potential of 27.5 $keV$.

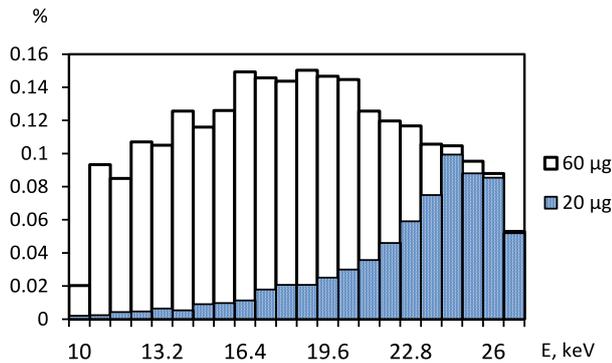

FIG. 6. Energy spectra comparison for the percentages of the backscattered protons. Dotted bars represent the detector with gold layer of 20 $\mu g/cm^2$ and empty bars represent detector with 60 $\mu g/cm^2$. The acceleration energy is 27.5 $keV$ in both cases.

In a thick gold layer, protons lose more energy just as expected. In addition, for a thicker gold layer, the threshold energy used is also larger. That is, protons fly out of the thick layer with less energy and they need more energy to break through the dead layer on the second attempt and leave a signal. Such a difference in the energy spectra prompted us to carry out a more detailed modeling, which does not require the hypothesis of an equal probability of detecting backscattered protons.

Instead of the extrapolation, we simulated the trajectories of protons backscattered into the electromagnetic field taking into account the angle of the proton's escape.

For backscattered protons, there are two reasons that reduce their chance of being detected on the second attempt. The first one is that particles scattered backward have less energy due to losses in the dead layer during the first pass. The second one is geometric; the cross section of elastic scattering has such a form that most of the particles will have a large angle relative to the normal. Therefore, the effective length of the dead layer increases and for some protons it is even possible to fly out of the detector radius, since the Larmor radius of motion of a proton in a field depends on the transverse velocity.

In our modeling, we have tried to take into account as many parameters as possible, including some parameters that were not taken into account by the authors of [1]. The distribution of protons by angles and energies at the entrance to the dead layer of the detector was simulated. The spectrum of protons in β-decay was also taken into account. The initial position of the proton was modeled implying the radial distribution of neutrons inside the trap which is shown in Fig. 5. For the backscattered protons, the energy and angle at which they fall back into the detector were determined. Also, we simulated both the passage through the dead layer and ionization in the active layer at the same time. This made it possible to more accurately estimate the energy loss in the dead layer and the probability of backscattering in the active layer, before the release of the amount of energy required for registration more accurately.

For each of the 9 energy+layer combinations that were used in measurements, we performed the simulation in several steps. At the first step, the passage of a sample of $3 \cdot 10^5$ protons through the dead layer was simulated. The protons were divided into registered, backscattered in a dead or active layer with the possibility of registration and lost ones. Losses included protons, either leaving too much energy in the dead layer and thus having energy below the threshold at the entrance to the active layer, or scattered back and having energy below the threshold value at the exit from the dead layer into the electromagnetic field.

At the second step, all the backscattered protons for which there is still a chance to be registered received new parameters of the entrance to the dead layer, and the simulation was performed for them again. At this stage, we checked that the protons remain within the detector. For the protons returning to the detector, the simulation of the passage through the dead layer was performed once more.

If a sufficiently large number of backscattered protons formed again, then the third step was performed, and so on. The procedure was terminated when the remaining fraction of backscattered protons after the next step became statistically insignificant.

As a result, instead of assuming the same probability of detecting a backscattered proton, we obtain for each layer+energy combination its own correction for lost protons. The simulation results are shown in Table II.

Table II For each layer+energy pair, the total loss values are given calculated both ways: directly and by extrapolation.

| Dead layer | E, keV | Total losses, Direct calculation, % | Total losses from the extrapolation, % |
|---|---|---|---|
| 20 | 27.5 | 0.045 | 0.497 |
| 20 | 30 | 0.047 | 0.434 |
| 20 | 32.5 | 0.037 | 0.357 |
| 60 | 27.5 | 2.72 | 2.56 |
| 60 | 32.5 | 1.34 | 1.76 |
| 0 | 30 | 0.465 | 0.344 |
| 0 | 32.5 | 0.339 | 0.286 |
| 40 | 27.5 | 0.533 | 1.24 |
| 0 | 27.5 | 0.614 | 0.445 |

The obtained result clearly demonstrates that the hypothesis of the same loss probability among backscattered protons is inapplicable. Therefore, the linear extrapolation of the results to the final value of the neutron lifetime based on it is also erroneous (unreliable).

To calculate the value of the neutron lifetime implying corrections we have obtained, one should calculate the corrected values and average them taking into account statistical errors. The measured and corrected values are presented in Table III.

Table III Measured and corrected neutron lifetime values for different energy+layer combinations.

| Dead layer | E, keV | Measured time, $\tau_{mes}$ | Corrected values $\tau_n$ |
|---|---|---|---|
| 20 | 27.5 | 892.4/884.1 | 892.0/883.7 |
| 20 | 30 | 885.9/889.1 | 885.5/888.7 |
| 20 | 32.5 | 889.3/891.8/892.3 | 889.0/891.5/892.0 |
| 60 | 27.5 | 909.9 | 885.2 |
| 60 | 32.5 | 901.1 | 889.0 |
| 0 | 30 | 888.0 | 883.9 |
| 0 | 32.5 | 890.7 | 887.7 |
| 40 | 27.5 | 899.0 | 894.2 |
| 0 | 27.5 | 888.5 | 883.0 |

The neutron lifetime obtained by us based on the data given in [1] is $\tau_n = 888.2 \pm 0.8\ s$. Fig. 7 shows two approaches to the estimation of the detector dead layer loss correction. For convenience, hereinafter, the corrected points are assigned $x$ coordinates corresponding to the points on the extrapolation. It should be remembered that these coordinates for the corrected points have no physical meaning and the result is obtained by averaging.

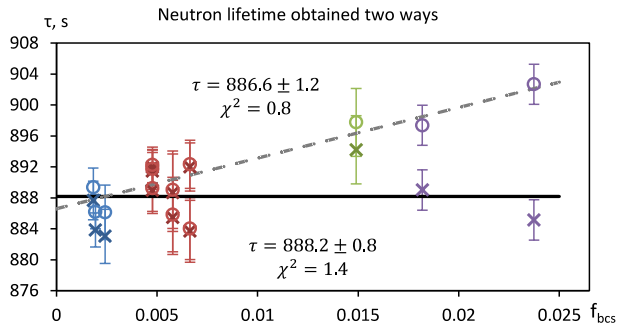

FIG. 7 A comparison of the methods of extrapolation (original points represented with circles and dashed line) and averaging (crosses and solid line), taking into account full corrections for losses in the dead layer.

This simulation can be refined even further by introducing more parameters of the experimental setup. For example, we used idealized configurations of electromagnetic fields; we do not know the exact values. In this work, our goal was not as much to obtain a new result, but to show that the modeling and its analysis given by the authors of [1] is, at least, insufficient and requires some clarification.

### III. RESIDUAL GAS AS A POSSIBLE SOURCE OF SYSTEMATIC ERROR

We have considered issues related to the motion and registration of protons after leaving the electromagnetic trap. The second part of the possible corrections is associated with the motion of protons inside the trap and the influence on this motion of the residual gas in the vacuum system. This problem was first discussed in [18], but we want to expand the research presented there. As in the previous section, we consider the configuration of the experimental setup from [1].

In this setup a vacuum system consists of a proton trap, a proton detector and a neutron monitor. Two ion pumps are located on the opposite sides relatively to the trap and they were used to pump out the vacuum system. Importantly, the proton detector and the neutron flux monitor had higher temperature than the proton trap during the measurements. Therefore, the residual gas pressure was measured in the warmer part of the vacuum system. Protons mostly reside in the proton trap; therefore the main parameter of our estimations is the concentration of the residual gas particles in the proton trap which was not directly measured in the experiment.

Essential parameters of the residual gas which directly affect the proton losses are composition and concentration. Direct measurements of the residual gas pressure were performed by the ionization vacuum gauge. The measured electrical current corresponded to the pressure of about $10^{-9}$ mbar. This quantity we will consider as a starting point in our analysis. In order to make a model of composition of the residual gas we use averaged data about the ion pumping [19]. Taking into account concentration and composition in the "warm" part of the vacuum system we estimate concentrations of gases inside the trap.

First and foremost, the evaluation relies on the temperature within the trap. At $10^{-9}$ mbar the mean free path of particles far exceeds the size of the vacuum system, therefore, molecular trajectories are determined by hits with the walls and any interaction between gas molecules is negligible. Therefore, the particle dynamics depends on the vacuum system geometrical configuration and the temperature of the walls. The proton trap can be considered as a long vessel with a small inlet. Gas molecules, which enter the trap, remain in it for a long time and reach thermodynamic equilibrium with the walls.

Consider a following model: a vessel with cold walls and a small inlet is located inside a closed vessel with warm walls and a small gas concentration inside. The initial gas concentration in the inner vessel is negligible. Gas from the outer vessel enters the inner one and after a few hits with the walls cools down to the inner vessel temperature, hence gas in the inner vessel has the same temperature as the walls. In the equilibrium state the amount of gas in the inner vessel is constant, but the warm gas enters it and the cold gas leaves. Hence, in the equilibrium state warm and cold fluxes at the inner vessel inlet have to be the same: $n_1 v_1 = n_2 v_2$ and $\frac{n_1}{n_2} = \frac{\sqrt{T_2}}{\sqrt{T_1}}$.

That model can be applied to the proton trap and for now neglecting freeze-out of molecules onto the trap surface we consider the upper bound estimation for the concentration of residual gas to be:

$$n_2 = \frac{P}{k\sqrt{T_1 T_2}} \quad (4)$$

where $P$ is pressure in warm part of the vacuum system, which is measured by the ion pump, $k$ is Boltzmann constant, $T_1$ and $T_2$ are the temperatures in the warm and cold parts. The warm part we consider to be at room temperature $300K$, and the cold part – at $4K$ which corresponds to the cooling of superconducting solenoid with liquid helium. With $P = 10^{-9}$ mbar we obtain the concentration in the cold part $n = 2.1 \times 10^8$ particles/cm³. If we consider the cold part temperature to be $10K$ or $20K$ we obtain the concentrations $1.3 \times 10^8$ particles/cm³ and $n = 9.3 \times 10^7$ particles/cm³.

Besides the total concentration, the calculations of the proton losses also require partial concentrations of the gases in the trap. We rely on the available data of the ion pumping to estimate the concentrations in the warm part of the trap. The composition of the residual gas depends on how the walls were prepared to the pumping. The high vacuum bake out is especially important because it strongly suppresses the amount of water in the residual gas. Most surfaces of the vacuum system were baked, but there were few exceptions, for example, the bore of the superconducting solenoid. During the experiment the direct measurements of the partial pressures of the gases were not performed. For that reason, in our analysis we considered the pumping with the various surface preparation techniques and obtain a range of possible losses.

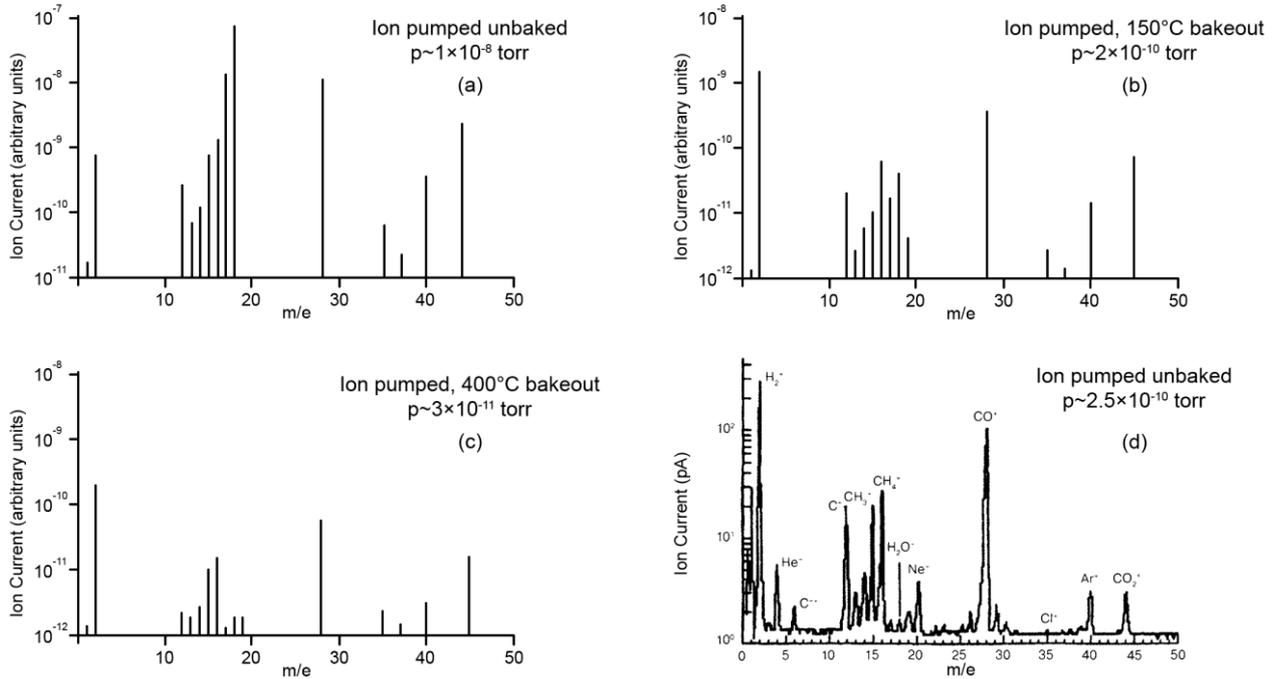

FIG. 8 The typical mass-spectra of the residual gas after the pumping out with the ion pump: a –no baking out, b – baking out at 150°C, c –baking out at 400, d – typical spectrum with the identified lines. Data taken from [19].

If a surface was not baked at the high vacuum then the main residual gas is the water vapor as shown in Fig. 8.a. If a surface was baked at 150°C then the concentration of water significantly decreases and for the most part the residual gas consists of hydrogen molecules with addition of the $CO$ molecules as shown in Fig. 8.b. With 400°C bake out the water vapor concentration becomes even smaller, but relative parts of $CO$, $CO_2$ and $CH_4$ are increased as shown in Fig. 8.c.

In all these cases we are interested in the relative concentrations, because the total concentration is determined by the temperature of the proton trap and the pressure of the residual gas in the warm part of the trap.

In Fig. 8 the data on the residual gases concentrations are presented in the form of the electrical currents in the measuring system. To extract gas concentrations from the currents one needs to use the coefficients which describe the effectiveness of registration for each gas and the data on fragmentation for each gas, because most molecular gases create several lines in the mass-spectrum. For example, $CO_2$ molecules create signals at $m/e$ equal to 44, 28, 16 and 12 corresponding to ions $CO_2^+$, $CO^+$, $O^+$, $C^+$. The data required to extract concentrations from the currents were taken in ref. [19]. Table IV shows the relative concentrations of the gases which have the biggest parts in the residual gas. The obtained spectra can be used to approximate composition of the residual gas in the proton trap in our calculations. This approximation can be used if adsorption in the trap can be neglected. At the very least, it can be used to obtain an upper limit of the proton losses.

Using concentration and composition of the residual gas one can perform qualitative and quantitative estimations of how it affects the proton losses. Two processes can contribute to proton losses: charge exchange and elastic scattering interaction of protons with the residual gas molecules.

Table IV Composition (%) of the residual gas in the vacuum system pumped out by the ion pump with various surface preparation techniques.

| Conditions\Gas | $H_2$ | $CH_4$ | $H_2O$ | $CO$ | $Ar$ | $CO_2$ |
|---|---|---|---|---|---|---|
| Unbaked $p \sim 10^{-8}$ torr | 0.4 | 1.5 | 88.7 | 6.5 | 0.4 | 2.2 |
| Bake out at 150 °C $p \sim 2 \cdot 10^{-10}$ torr | 63.0 | 4.6 | 4.7 | 18.9 | 1.5 | 5.6 |
| Bake out at 400 °C $p \sim 3 \cdot 10^{-11}$ torr | 52.5 | 11.8 | 1.5 | 22.3 | 1.6 | 7.4 |

## IV. THE PROTON LOSSES IN THE ELASTIC SCATTERING

At first we consider elastic scattering of protons on the residual gas on the measurement result.

The electrostatic system of the experimental setup [1] consists of a cylindrical cavity with a solenoid, which creates the static magnetic field, and a system of electrodes. This system can be divided in two parts 1)"trap electrodes" or the grounded electrodes with zero potential, 2)"mirror" electrodes – the electrodes situated at the ends of the trap and which determine an area where the axial component of the velocity $v_z$ changes its sign (see Fig. 1).

There are three modes of the trap operation: 1) trapping protons, 2) counting protons, 3) cleaning the trap. β-decay protons created in the trap move between the mirror electrodes in the trapping mode, because the maximal kinetic energy of the decay protons is less than the potential barrier between the grounding and the mirror electrodes. In the counting mode the electric field is applied in the trap to guide the protons to the detector. Finally, in the cleaning mode the electric field pushes all charged particles out of the trap and prepares it to the next measurement cycle. Further we analyze the proton dynamics in details to estimate possible losses in the trapping mode.

In the general case, a proton created in the trap has a complicated trajectory limited in space (see Fig. 9.) with oscillations of three types: 1) axial – along the axis of the solenoid which is caused by the gap in potentials of electrodes, 2) azimuthal – the rotation around the solenoid axis, 3) cyclotron – the rotation around the guiding center. The guiding center moves along the magnetic field lines and drifts in the electric field with a velocity $\boldsymbol{v_E} = (\boldsymbol{E} \times \boldsymbol{B})/B^2$ orthogonal to the magnetic field lines. Each oscillation has its own period.

In the experimental setup the trap radius, the magnetic field strength and the neutron beam width are tuned in a way that the β-decay protons always reach the detector. However, losses caused by the elastic scattering were not taken into account. To estimate these losses we carried out special simulation of the proton motion within the trap considering the elastic scattering at the residual gas. In this simulation we used the elastic scattering cross-sections of protons at the hydrogen, because according to our analysis of corrections for charge exchange process, the molecular hydrogen is the main part in the residual gas inside the proton trap.

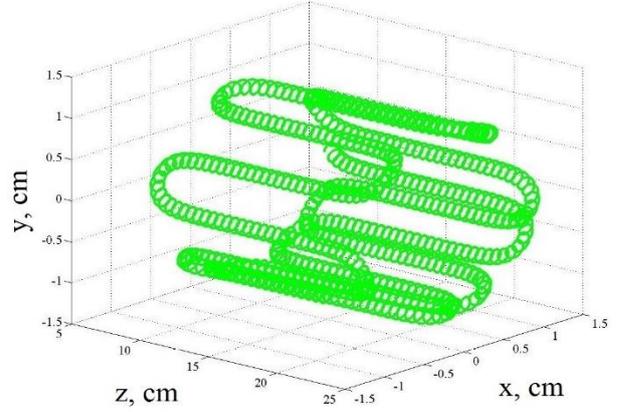

FIG. 9. The trajectory of a proton with the initial kinetic energy 700eV in the trap. Oz – the solenoid axis.

In our simulation we assumed that the elastic scattering at an atom occurs at some point of a proton trajectory and as a result the proton velocity changes its direction with scattering angle $\theta$. Using the differential cross-section of the elastic scattering and the energy conservation law for each pre-hit proton velocity $v_0$ the probability distribution function of the scattering angles $\theta$ was calculated. The next step was to calculate how post-hit proton velocity depends on its pre-hit velocity and the scattering angle, and using the obtained post-hit velocity we calculate a new mean time between the hits $\tau$. The calculated functions were represented in a form of tables and these tables were used in the simulation of the proton trajectories with the elastic scattering at the residual gas.

From the simulation results one can conclude that if there is high concentration of the residual gas in the system and if the initial radial coordinate is close to the detectors effective radius 9 mm, after 5-7 scatterings in angles close to 90 degrees the cyclotron radius of the proton can exceed the effective radius of the detector.

We are interested not only in the correction for the elastic scattering at a specific pressure by itself but its potential range also. Therefore, we investigated how an increase in the concentration of the residual gas affects the trajectories of protons in the trap. In Fig. 10 one can see how a projection of a proton trajectory on the plane orthogonal to the solenoid axis changes along with increasing of residual gas concentration. At low concentrations the trajectory is projected on a set of points located between two concentric circles and hence changes of radial coordinate are limited. Starting from a certain concentration frequency of the hits increases to values at which protons can be lost due to unbalanced fluctuations of the radial coordinate (see Fig. 10.b). Further increase of the concentration results in more curved proton trajectories which are not limited by the cylindrical surfaces (see Fig. 10.c).

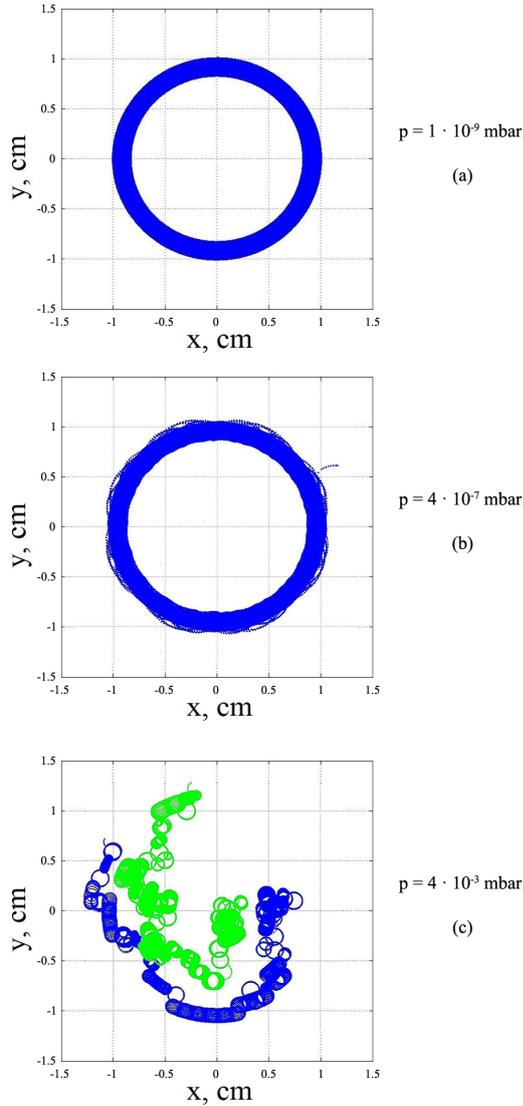

FIG. 10. Projections of proton trajectories on a plane orthogonal to solenoid axis obtained in simulations of elastic scattering in proton trap for various concentrations of Oxygen atoms: a) small concentration, b) medium concentration, c) high concentration.

The scale of the residual gas pressure necessary for each scenario is illustrated in Fig. 10. From these results, it becomes clear that even the maximum possible pressure of the residual gas inside the trap according to formula (4) is insufficient to significantly affect the trajectories of protons. It can be concluded that elastic scattering on the residual gas introduces a negligible correction compared to the systematic error caused by the proton charge exchange process.

## V. CHARGE EXCHANGE PROCESS AS THE SOURCE OF PROTON LOSSES

Here we consider the charge exchange process between protons and residual gas atoms: $H^+ + A \rightarrow H + A^+$. The charge exchange cross-sections of protons with velocities much less than the classical electron velocity in an atom have rather big values for many gases and significantly exceed the elastic scattering cross-sections. Therefore, this process may cause systematic error. The $\beta$-decay protons do not get an additional acceleration in the proton trap so their energies are less than 750 eV and the velocities are more than an order of magnitude less than the classical velocity of an electron in an atom.

We have previously discussed possible residual gas compositions. Now we want to consider the scale of its influence on uncertainties. The charge exchange process leads to the appearance of a hydrogen atom, which escapes from the trap, and a charged molecular ion continues to move inside the magnetic trap like the proton would. And after a storage period it is directed to the detector. Therefore, the ion can create a signal in the detector above the set threshold, which means that it is necessary to take into account the probability of ion loss when passing through the dead layer of the detector.

For each gas, it is possible to calculate the probability of a proton loss as a result of the proton charge exchange process on this gas. This probability consists of three factors: the first $K_i$ is the loss coefficient, which depends on the charge exchange cross section, the energy, and the proton path length in the trap. Essentially, this is the probability of the charge exchange per particle per unit volume. The second factor is the concentration of particles of a given gas per unit volume. The third is the probability that the accelerated molecular ion will not be registered by the detector and the charge exchange process will actually lead to the signal loss. In the previous part, we gave a list of gases that are most likely to be present in a vacuum system. For each of these gases, as a first step we calculate the coefficient $K_i$ using an expression:

$$K_i = \int \sigma(E) v(E) t_m P(E) dE \quad (5)$$

where $\sigma(E)$ is the charge exchange cross section of this gas, $v(E)$ is a proton velocity at a given energy, $t_m$ is the mean time of the proton storage in the trap, $P(E)$ is the normalized $\beta$-decay protons spectrum. The coefficient $K_i$ has the dimension of $cm^3$.

In our calculations we used the proton charge exchange cross-sections available from the various published sources. The list of the considered gases, the corresponding loss coefficients, and the sources of the data about the cross-sections are presented in Table V. The corresponding charge exchange cross-sections are shown in Fig. 11. Since there are no helium peaks in the mass-spectra in Fig. 8 and the charge exchange cross section for the helium atoms about four orders of magnitude smaller than of the presented gases it is omitted from Fig. 11 and the tables.

Table V The loss coefficients of the protons for the gases having the largest parts in the residual gas.

| Target | Loss coefficient $cm^3$ | Data source |
|---|---|---|
| $H_2$ | $2.426 \cdot 10^{-11}$ | [20] |
| $CH_4$ | $3.332 \cdot 10^{-10}$ | [21] |
| $H_2O$ | $3.050 \cdot 10^{-10}$ | [22] |
| $CO$ | $1.896 \cdot 10^{-10}$ | [22, 23] |
| $Ar$ | $4.872 \cdot 10^{-11}$ | [24] |
| $CO_2$ | $2.617 \cdot 10^{-10}$ | [22] |

In general, the accuracy of the measured cross-sections is about 10-20% depending on the gas. For most of the gases one can find an analytical approximating expression which fits the data with accuracy of the same level as the experimental accuracy. These expressions were used in our calculations. For the other gases we used the interpolations of the existing experimental data. In the energy region below 100 $eV$ for some gases experimental data do not exist or have the higher uncertainties, but, fortunately, that region contributes very little to the calculated coefficients. Only 8% of the protons have energies below 100 $eV$ and such protons have the shorter paths in the trap and hence the low probabilities of losses due to the charge exchange process.

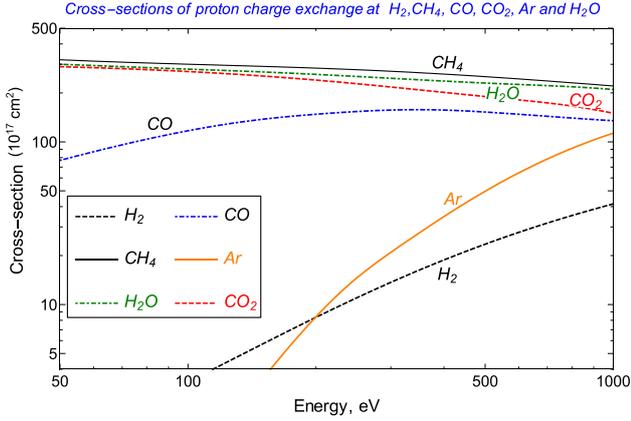

FIG. 11. The charge exchange cross sections of the molecules which have the largest parts in the residual gas.

In our analysis we do not consider the further interactions of the molecular ions created in the charge exchange process with the residual gas molecules, since the probability of such interactions are proportional to the squared loss coefficient from Table V. After the storage period the ions (and the charged molecules) along with the protons are accelerated in the electric field and reach the detector.

The charge exchange process results in a signal loss if an ion would be absorbed or scattered back in the detector dead layer or if the signal of the ion is under the detector's discriminator threshold. Therefore, the second step is to calculate the probability of loss of a molecular ion during detection, which we denote as $L_i$. Earlier, we showed that the detection probability for the protons strongly depends on the type of the dead layer and the magnitude of the accelerating potential. Therefore, different energy+layer combinations must be considered separately here as well. The analysis of that systematics requires modeling the passage of ions through the dead layer for different combinations. In our simulations, we made several assumptions. The initial momenta of molecular ions can be neglected and it can be assumed that the ion energy at the entrance to the dead layer is equal to the accelerating potential. The energy is evenly distributed among the individual atoms according to their relative masses. For the molecular ions, we do not consider re-passage through the dead layer, when they are backscattered, since it can almost always be neglected. The average travel distance for an ion in a material is very different for different ions, even at the same energy. In general, the greater the mass and charge of an ion, the less distance it will travel. Fig. 12 represents that difference of the stopping ranges for the hydrogen and oxygen ions. Left edge of the plot represents entrance to the detector and right edge is 500 nm inside. Therefore, for each gas and for each layer there will be a different probability of losses. For all gases listed in Table IV and for each energy+layer combination, the fraction of ions not passing through the dead layer was calculated. The results are shown in Table VI. As like as for protons, the simulation was carried out using the SRIM2013 program. Fig. 13 illustrates the results shown in Table VI for a layer of $20\,\mu g/cm^2$ at the accelerating potential of 30 $keV$. Shown here is the calculated detector response for various gases, taking the relative density into account. For modeling, the total concentration corresponding to the temperature 4K equation (4) and the relative concentrations corresponding to the second row of Table IV were taken.

This calculation confirms that different types of layers must be considered separately and for each one must calculate its own correction. The coefficient $L_i$ in the correction acts linearly, which means that the correction for the charge exchange on water molecules will differ for the thin layer of gold and the layer of silicon oxide by 4-5 times.

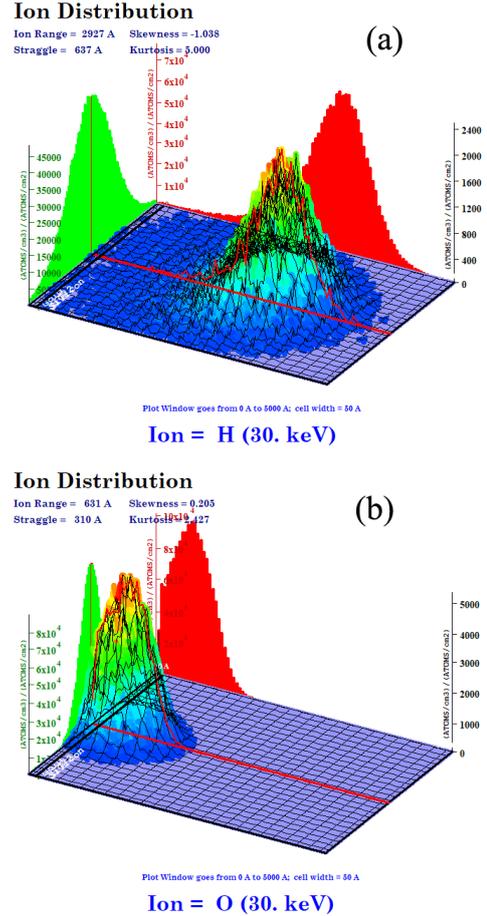

FIG. 12 Ion distribution plots for hydrogen (a) and oxygen (b) with initial energy of 30 keV passing through the 100Å layer of gold than the 25Å layer of silicon oxide and finally stopping in the pure silicon.

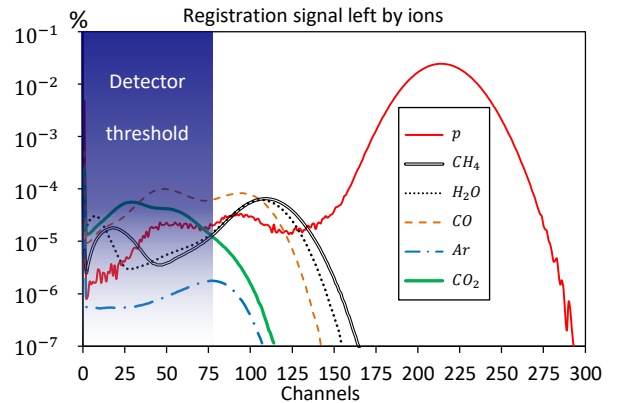

FIG. 13 The simulation of signals of molecular ions for the molecules having the largest part in the residual gas.

Table VI. The Loss probability $L_i$ of accelerated molecular ions in the detector dead layer

| Gas | Au 20 $\frac{\mu g}{cm^2}$ 27.5 keV | Au 20 $\frac{\mu g}{cm^2}$ 30 keV | Au 20 $\frac{\mu g}{cm^2}$ 32.5 keV | Au 40 $\frac{\mu g}{cm^2}$ 27.5 keV | Au 60 $\frac{\mu g}{cm^2}$ 27.5 keV | Au 60 $\frac{\mu g}{cm^2}$ 32.5 keV | $SiO_2$ 27.5 keV | $SiO_2$ 30 keV | $SiO_2$ 32.5 keV |
|---|---|---|---|---|---|---|---|---|---|
| $H_2$ | 0.001 | 0.001 | 0.001 | 0.032 | 0.213 | 0.061 | 0.086 | 0.056 | 0.038 |
| $CH_4$ | 0.216 | 0.199 | 0.167 | 0.843 | 0.998 | 0.957 | 1 | 1 | 1 |
| $H_2O$ | 0.26 | 0.263 | 0.206 | 0.826 | 0.995 | 0.945 | 1 | 1 | 1 |
| $CO$ | 0.557 | 0.547 | 0.478 | 0.994 | 1 | 1 | 1 | 1 | 1 |
| $Ar$ | 0.66 | 0.66 | 0.585 | 0.995 | 1 | 1 | 1 | 1 | 1 |
| $CO_2$ | 0.88 | 0.88 | 81.5 | 1 | 1 | 1 | 1 | 1 | 1 |

To obtain the final corrections for the various measured neutron lifetime values, it remains to estimate the concentration and composition of the residual gas. The total pressure of the residual gas in the warm part of the setup in the experiment under consideration was measured to be $10^{-9}$ mbar. We will use this value for further analysis. In Table IV, in addition to the relative concentrations, the absolute values of the gas pressure are also given, which are typical for the different options of baking the experimental setups. The closest pressure value to that measured in the experiment is one corresponding to the case where the setup is heated to 150 C — $p \sim 2 \cdot 10^{-10}$ torr. However, this limiting pressure is about 5 times better than the experimentally measured value, but it can be explained by the fact that not all parts of the vacuum system were baked before the experiment. Taking into account these data, for the further analysis and modeling we will take the relative concentrations given in the second row of Table IV, at the total pressure in the warm part equal to $10^{-9}$ mbar.

For each energy+layer combination the fraction of the protons that is lost due to the charge exchange process at a gas is a product of the concentration, the loss coefficient, and the loss probability of that gas. To obtain the total proton losses we should sum the losses over all gases:

$$\delta_{loss} = \sum_i n_i \cdot L_i \cdot K_i \qquad (6)$$

Using the equation (6), the coefficients from Table V and VII, and the gas concentrations from Table IV we calculated the losses for all energy+layer combinations. The total gas concentration inside the trap was calculated using formula (4) for different possible temperatures. In this case, the temperature inside the trap equal to 300 K corresponds to the pressure inside the trap equal to the pressure in the warm part of the vacuum system.

Fig. 14 shows two sets of points: excluding the residual gas correction (see Table III) and taking into account the residual gas at a pressure corresponding to the temperature in the trap 4K.

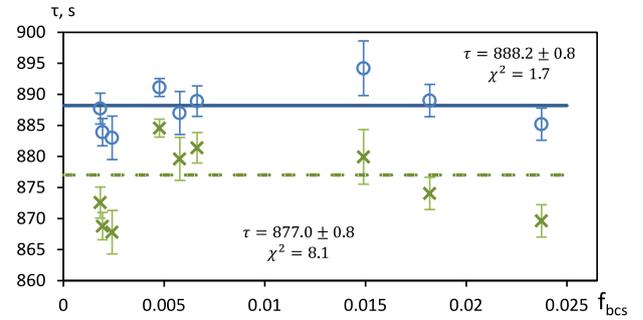

FIG. 14. Corrections for the residual gas at the concentration corresponding to the temperature of 4K. Circles – values corrected only for the dead layer of the detector. Crosses – values corrected for the dead layer and the charge exchange on the residual gas. The temperature inside the trap is 4K.

The difference between the average values is 11 seconds, but in this case, not only the absolute value of the possible correction is important, but also the difference between the corrections for different types of layers. Table VII shows the corrections in seconds calculated for each layer for 5 temperatures.

Table VII. Corrections for the total gas composition. Each cell contains value in seconds. Zero gold thickness corresponds to the silicon oxide dead layer.

| | Thickness of gold ($\mu g/cm^2$) / Acceleration potential ($keV$) | | | | | | | | |
|---|---|---|---|---|---|---|---|---|---|
| T, K | 20/27.5 | 20/30 | 20/32.5 | 40/27.5 | 60/27.5 | 60/32.5 | 0/27.5 | 0/30 | 0/32.5 |
| 4 | 7.51 | 7.39 | 6.53 | 14.26 | 15.56 | 14.95 | 15.18 | 15.11 | 15.13 |
| 10 | 4.75 | 4.67 | 4.13 | 9.02 | 9.84 | 9.45 | 9.6 | 9.56 | 9.57 |
| 20 | 3.36 | 3.3 | 2.92 | 6.38 | 6.96 | 6.68 | 6.79 | 6.76 | 6.77 |
| 100 | 1.5 | 1.48 | 1.31 | 2.85 | 3.11 | 2.99 | 3.04 | 3.02 | 3.03 |
| 300 | 0.87 | 0.85 | 0.75 | 1.65 | 1.8 | 1.73 | 1.75 | 1.75 | 1.75 |

Two conclusions can be drawn from the obtained results. First, for different configurations of the dead layer, the residual gas corrections are substantially different. This is yet another argument against using the linear extrapolation of the data obtained with the different thicknesses of the dead layer. For each configuration of the dead layer and each value of the accelerating potential, it is necessary to separately calculate both the correction for the loss of protons in the dead layer and the correction for the loss of protons as the result of charge exchange on the residual gas.

Second conclusion — for low temperatures, the calculated corrections are very large, but at the same time, this difference does not manifest itself in the experimental data. The calculated corrections significantly exceed the statistical errors, especially for the thick layers of gold and for the layer of silicon oxide. This is expressed in a too large value of $\chi^2 = 8.1$.

If the considered composition of the residual gas was really the case in the experiment, then the measurements carried out with the detector covered with the dead layer of silicon oxide would have given values much larger than the measurements carried out with the detectors covered with the thin layer of gold. But this effect is not seen in the experimental data.

There can be two explanations for this: the upper limit on the particles density calculated by formula (4) is not applicable for the proton trap, or the composition of the residual gas differs from the values given in Table IV. Actually, most probable that it is both these scenarios and the observed overcorrection shown in Table VII can be naturally explained by freeze-out of the residual gas components onto the trap surface.

As shown in Table IV, the gases, which are in the most part responsible for the proton losses in the trap, are $H_2$, $CH_4$, $CO$, $CO_2$, $Ar$ and $H_2O$. The concentration of water significantly decreases after the baking and it also can be frozen out at the low temperatures. However, $H_2$, $CH_4$, $CO$, $CO_2$ have the vapor pressure $1 \cdot 10^{-8}$ torr at temperatures 4К, 32К, 27К and 76К correspondingly [25]. These gases also condense on the surfaces at higher temperatures, but above the surfaces with the listed temperatures the pressure would be $1 \cdot 10^{-8}$ torr which is even an order of magnitude higher than the pressure used in the calculations.

Freezing-out of various gases on the surface of the proton trap can be taken into account. When the temperature drops, water will condense on the surface first, followed by carbon dioxide. If one removes these gases from the mixture, then the corrections reduce significantly. Corresponding corrections are given in Table VIII. Fig. 15 illustrates the first and third rows of Table VIII. The first row corresponds to the maximum possible corrections for the freezed-out water and $CO_2$, and the third corresponds to the temperatures close to the freezing temperatures of $CH_4$ and $CO$.

The largest contribution to the correction comes from the residual carbon monoxide, which, moreover, freezes out rather poorly inside a cold trap and its effect cannot be neglected until there are direct studies of the composition and concentration of the residual gas inside the proton trap.

Table VIII. Corrections for $H_2O$ и $CO_2$ freeze-out.

|      | Thickness of gold ($\mu g/cm^2$) / Acceleration potential ($keV$) | | | | | | | | |
|------|--------|--------|--------|--------|--------|--------|--------|--------|--------|
| T, K | 20/27.5 | 20/30 | 20/32.5 | 40/27.5 | 60/27.5 | 60/32.5 | 0/27.5 | 0/30 | 0/32.5 |
| 4    | 4.42   | 4.29   | 3.75   | 9.31   | 10.21  | 9.7    | 9.83   | 9.75   | 9.74   |
| 10   | 2.8    | 2.72   | 2.37   | 5.89   | 6.45   | 6.14   | 6.22   | 6.17   | 6.16   |
| 20   | 1.98   | 1.92   | 1.68   | 4.16   | 4.56   | 4.34   | 4.4    | 4.36   | 4.36   |

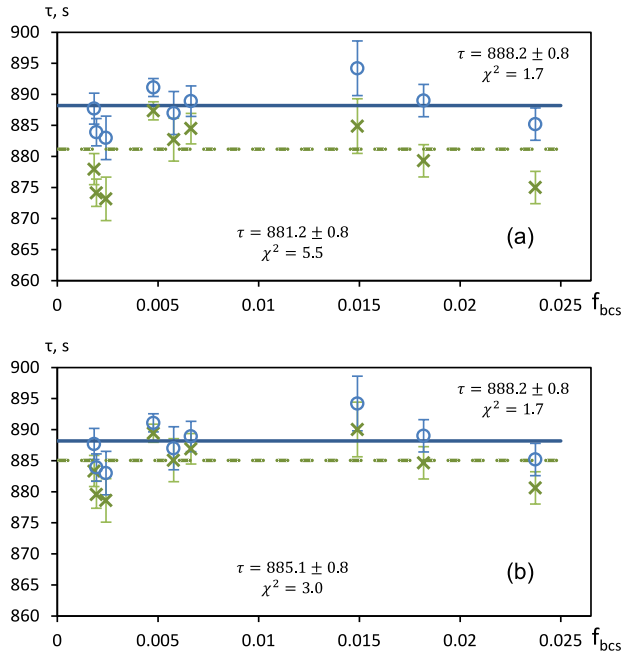

FIG. 15 Circles – the values corrected only for the dead layer of the detector. Crosses – values corrected for the dead layer and the charge exchange process on the residual gas considering that $H_2O$ and $CO_2$ freeze-out. Temperature inside the trap is 4K (a) or 20K (b).

If the temperature inside the proton trap lies in the range of 20-30K, then corrections of the order given in the third row of Table VIII should be applied to the measured values. These errors are already comparable with the statistical error of the experiment. However, note that they are still larger than the corrections calculated for the case of all gases at the temperature of 300K.

If the real temperature inside the trap is even lower and it can be assumed that all gases except molecular hydrogen are completely frozen-out on the trap walls, then the corrections for the charge exchange become imperceptible against the background of the statistical error.

In this case, the corrections generated by the residual gas are only $0.13\ s$, which is shown in Table IX and it is much less than the statistical error of the experiment.

Our calculations show that the measurements with different types of detectors can be used to assess the effect of residual gas. The correction for residual molecular hydrogen for detectors with the thick and the thin layers of gold can differ by a factor of 8, although it has a very low value, therefore, to use this difference, a very high statistical accuracy must be obtained.

Table IX Corrections for the case when only $H_2$ remains as a residual gas.

|      | Thickness of gold ($\mu g/cm^2$) / Acceleration potential ($keV$) | | | | | | | | |
|------|--------|--------|--------|--------|--------|--------|--------|--------|--------|
| T, K | 20/27.5 | 20/30 | 20/32.5 | 40/27.5 | 60/27.5 | 60/32.5 | 0/27.5 | 0/30 | 0/32.5 |
| 4    | 0.003  | 0.003  | 0.003  | 0.091  | 0.603  | 0.173  | 0.243  | 0.158  | 0.108  |
| 10   | 0.002  | 0.002  | 0.002  | 0.058  | 0.381  | 0.11   | 0.154  | 0.1    | 0.068  |

Here we should mention another possible source of the systematic error. Besides the energy release in the detector there is another parameter that can affect the signal. The molecular ion produced in the charge exchange process is heavier than the proton and hence it hits the detector later. The fact that initially ions are very slow in comparison with the protons further increases the time delay.

If the data collecting system records data for a limited time period then the signals produced by the molecular ions that managed to pass the dead layer still can be cut off. Therefore the limited data collecting period can further affect the loss coefficients.

The Ref [1] provides the information of data collecting period of each measurement cycle. It longs for about $30\mu s$. The time spectrum of the signal demonstrates that such period is sufficient to collect all protons from the trap. But to include timing cut in the loss probability of molecular ions one has to estimate how long will it take for the heavier ion with small initial speed to reach the detector. The accurate simulation of this effect requires full description of the setup including the details of its geometric configuration and applied electromagnetic fields.

Since we had only limited information we made a rough estimation that can reveal only the main features of taking the time spectrum into consideration.

The Fig 6. in Ref [1] demonstrates the time spectrum of the proton signal. We added the estimated spectrum of $H_2$ signals and the result is shown in Fig. 9. Heavier gases fall out of that diagram and for them we can consider the probabilities of producing proton like signals to be zero, if we take into account time cutoff.

The molecules heavier than $H_2$ already have high loss probabilities from the dead layer. If we assume that some of that gases do not freeze out and include for them the timing cut than it only further increase the corrections obtained above for a relatively small value. But if we assume total freeze out of all gases except the $H_2$ then the time cut correction would dominate in loss probability. Therefore we would like to focus on the latter case and add the timing cut only for $H_2$ molecules.

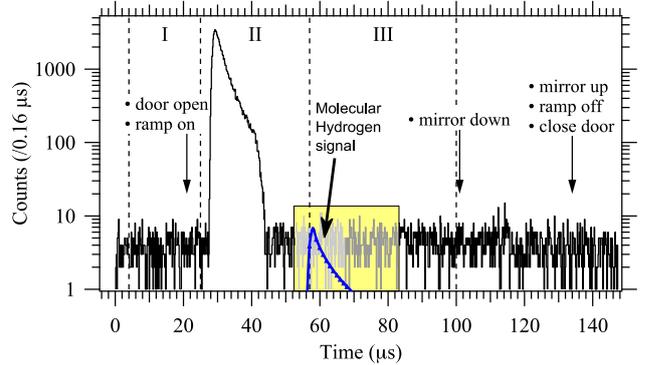

FIG. 16 Measured timing spectrum from ref. [1] with the molecular hydrogen response added.

In our estimation only 27% of the $H_2$ signal falls into region 2 and the rest lies in region 3. The amount of $H_2$ signals was estimated by taking into account charge exchange probability $K$ (Table V). We assume that signals which fall into region 2 are simply counted as protons. But the signals that fall into region 3 have even more potential to create systematic error than simple losses. Counts in region 3 are used to calculate the background and therefore subtracted from the counts in region 2. Therefore if the proton is lost in charge exchange process at the $H_2$ molecule and the signal falls into the third region then we should consider loss of about 5/3 of the signal because of the subtraction.

In Table X we present the corrections obtained for each dead layer after analysis of losses in the dead layer and losses connected with the time cutoff.

Table X Corrections for the case when only $H_2$ remains as a residual gas including the time cutoff.

| | Thickness of gold ($\mu g/cm^2$) / Acceleration potential ($keV$) | | | | | | | | |
|---|---|---|---|---|---|---|---|---|---|
| T, K | 20/27.5 | 20/30 | 20/32.5 | 40/27.5 | 60/27.5 | 60/32.5 | 0/27.5 | 0/30 | 0/32.5 |
| 4 | 3.46 | 3.45 | 3.47 | 3.48 | 3.41 | 3.47 | 3.40 | 3.42 | 3.44 |
| 10 | 2.19 | 2.18 | 2.19 | 2.20 | 2.15 | 2.19 | 2.15 | 2.16 | 218 |

The comparison with the correction listed in Table IX shows that in case of only molecular hydrogen in the trap the main part of the correction is due to timing cut. If we consider only molecular hydrogen as a residual gas but take into account both factors the losses in the dead layer and loses due to the timing cut then the total correction for neutron lifetime is 3.4 s, as shown in Fig. 17. That makes the $H_2$ concentration in the trap one of the most important parameters which is essential to accurately calculate the systematic uncertainty of the beam experiment.

As a result of our analysis we obtained several possible corrections. We did not set the task of calculating the final correction to the neutron lifetime measured in the experiment [1]. It was important for us to investigate the spectrum of possible systematic effects associated with the charge exchange process on the residual gas. As a result, we come to the conclusion that the residual gas can potentially make a significant contribution to the systematic error, but without direct measurements of its density and composition in the proton trap, it is not possible to obtain a sufficiently accurate value of the correction.

The second conclusion that can be drawn here is that a cold trap is effective only if there is absolute confidence that all residual gas, except helium, freezes out on the walls, otherwise the proton trap becomes a cryogenic storage of residual gas, cooling it down to low temperatures and increasing its concentration.

For a warm trap, the correction for the residual gas turns out to be much smaller than in the case of a cold trap if the carbon monoxide and methane are not frozen out or if we consider the time cutoff for molecular hydrogen. Our calculations show that to obtain 1 s accuracy with a warm trap vacuum should be better than $2 \cdot 10^{-10}$ torr, which can be reached with a moderate baking at the temperature 150°C, but it is still almost an order of magnitude better vacuum than the one used in the experiment. A warm trap has to be separated from a cold superconducting solenoid which has the temperature 4K.

In this case, the efficiency of freeze-out effect can also be used in the experiment. The installation can consist of a warm proton trap and a cold surface, at which the most dangerous components of residual gas can freeze-out.

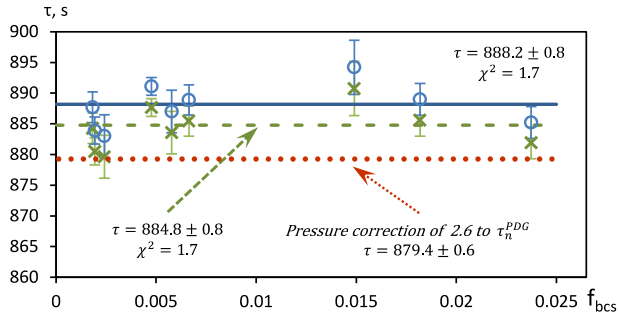

FIG. 17 Circles – the values corrected only for the dead layer of the detector. Crosses – values corrected for the dead layer and the charge exchange process on the $H_2$ molecules with taking the time cut into consideration. Dotted line which represents $\tau_n^{PDG}$ and the amount of possible pressure correction that might lead to that value are discussed further in this chapter.

In any case, carrying out an experiment with a proton trap requires monitoring of the parameters of the residual gas and numerical modeling of the systematic error introduced by its presence.

The charge exchange correction is directly proportional to the residual gas pressure. As shown in Fig. 17 even if the only residual gas in the trap is $H_2$ the correction can be significant. Comparing the obtained correction with the current value for the neutron lifetime of $\tau_N^{PDG} = 879.4 \pm 0.6\,s$ provided by the Particle Data Group we can deduce that if real residual gas pressure in the trap cited in the Ref. [1] as $10^{-9}\,mbar$ might differ by the factor of only 2.6 than it could completely resolve existing neutron anomaly. Therefore, including the correction caused by charge exchange of the protons at $H_2$ is the most probable solution of the neutron anomaly.

The method of accounting for proton losses by the linear extrapolation of the data obtained with different detectors is inapplicable neither for proton losses in the dead layer, nor for losses due to charge exchange process, since the effects of the dead layer are independent and must be considered separately. For each measured lifetime, it is necessary to calculate individual corrections for interaction with the residual gas and backscattering of protons, and average the corrected values.

To achieve the highest accuracy with a cold trap, it is necessary to choose the best temperature value at which the main residual gas components are frozen out on the trap walls and at the same time hydrogen and helium would not start to accumulate inside the trap.

## VI. CONCLUSIONS

We have investigated several possible systematic corrections in a beam-type experiment of measuring the neutron lifetime. In this study, we focused on the corrections associated with the registration of protons in the experiment [1], where the most accurate value of the neutron lifetime was obtained by the beam method. For the registration of protons three problems can be highlighted — the motion of protons, their registration, and the residual gas influence.

As a result of simulating the motion of protons in the electromagnetic field, we came to the conclusion that the width of the proton flux coming from the trap does not exceed the dimensions of the detector and the corresponding systematics does not appear in the experiment. Additional allowance for elastic scattering by the residual gas molecules also does not lead to the proton losses and systematic errors. The potential influence of the elastic scattering was investigated in details, and we determined the concentrations of the residual gas, at which its influence becomes noticeable and cannot be neglected, but these concentrations are significantly higher than the upper limit of the concentration of the residual gas in a vacuum part of the setup.

For the problem of registering protons with a semiconductor detector, the simulation was carried out that showed that the method of the linear extrapolation for determining the losses among backscattered protons in the dead layer is inapplicable. Instead, for each type of the dead layer, it is necessary to independently calculate the correction for the total loss associated with the passage of protons through the dead layer and the correction for backscattering in the active layer of the detector.

The charge exchange of protons on the residual gas in a proton trap can potentially make the largest contribution to the loss of protons. This effect essentially depends on the concentration and composition of the residual gas in the trap. The freezing out of gases on the cold walls of the trap do not solve that problem because even $H_2$ can produce the significant error and even might be the leading cause of the neutron lifetime anomaly. We investigated the spectrum of the possible corrections and showed that, depending on the amount of gas, the correction can vary from values exceeding the statistical error of the experiment, to values that are invisible against the background of the statistical errors. We draw two conclusions for this problem. First, the accurate accounting of the charge exchange in the residual gas requires direct measurements of the concentration and composition of the residual gas inside the proton trap. Secondly, the cold trap turns into a cryogenic pump, in which the concentration of the residual gas turns out to be higher than in the warm part. Therefore, in the future beam experiments and in current efforts to improve measurement accuracy, the possibility of creating a proton trap without cooling should be considered. At the same time, to pump out such a trap, in addition to the ion pump, one can also use a cryogenic pump, since it is more efficient for pumping out gases with large values of the charge exchange cross section.

To solve the problem of the neutron anomaly, it is necessary to carry out a new experiment to measure the neutron lifetime using the beam method. And such experiments are currently being developed in particular experiment of Hoogerheide et al [15] dedicated to investigation of systematic errors with renewed setup and with more intense neutron beam. We hope that the results of our research will be useful for our colleagues.

This work was supported by the Russian Foundation for Basic Research (Project № 19-02-00582-a).